\documentclass[acmsmall]{acmart}

\AtBeginDocument{%
  \providecommand\BibTeX{{%
    \normalfont B\kern-0.5em{\scshape i\kern-0.25em b}\kern-0.8em\TeX}}}

\usepackage{xcolor}
\usepackage{color,soul}
\usepackage{mwe}
\usepackage{xspace}
\usepackage{etoolbox}
\usepackage{enumitem}

\hypersetup{colorlinks=true,linkcolor=blue,urlcolor=blue}
\AtEndEnvironment{quoting}{\vspace{3\baselineskip}}
\AtEndEnvironment{quoting}{\vspace{3\baselineskip}}

\definecolor{orange}{RGB}{215,87,0}
\definecolor{red}{RGB}{220,0,0}
\definecolor{agreen}{RGB}{74, 198, 148}
\definecolor{purple}{RGB}{173, 141,174}
\definecolor{aqua}{RGB}{87, 180, 181}
\definecolor{RoyalBlue}{rgb}{0.36, 0.54, 0.66}
\colorlet{lightmintbg}{RoyalBlue!40}
\colorlet{lightermintbg}{RoyalBlue!50}
\colorlet{lightestmintbg}{RoyalBlue!60}

\newcommand{\tool}[0]{\textsc{Recast\xspace{}}}

\begin{document}

\setcopyright{acmlicensed}
\acmJournal{PACMHCI}
\acmYear{2021} \acmVolume{5} \acmNumber{CSCW1} \acmArticle{181} \acmMonth{4} \acmPrice{15.00}\acmDOI{10.1145/3449280}

\title[\tool{}]{\tool{}: Enabling User Recourse and Interpretability of Toxicity Detection Models with Interactive Visualization
}


\author{Austin P Wright}
    \affiliation{Georgia Institute of Technology, USA} 
    \email{apwright@gatech.edu} 
\author{Omar Shaikh}
    \affiliation{Georgia Institute of Technology, USA}
    \email{oshaikh@gatech.edu} 
\author{Haekyu Park}
    \affiliation{Georgia Institute of Technology, USA}
    \email{haekyu@gatech.edu} 
\author{Will Epperson}
    \affiliation{Georgia Institute of Technology, USA}
    \email{willepp@gatech.edu} 
\author{Muhammed Ahmed}
    \affiliation{Mailchimp, USA}
    \email{muhammed.ahmed@mailchimp.com}  
\author{Stephane Pinel}
    \affiliation{Mailchimp, USA}
    \email{stephane.pinel@mailchimp.com} 
\author{Duen Horng (Polo) Chau}
    \affiliation{Georgia Institute of Technology, USA}
    \email{polo@gatech.edu} 
\author{Diyi Yang}
    \affiliation{Georgia Institute of Technology, USA}
    \email{diyi.yang@cc.gatech.edu}

\renewcommand{\shortauthors}{Austin P. Wright et al.}


\begin{abstract}
With the widespread use of toxic language online, platforms are increasingly using automated systems that leverage advances in natural language processing to automatically flag and remove toxic comments. 
However, most automated systems---when detecting and moderating toxic language---do not provide feedback to their users, let alone provide an avenue of recourse for these users to make actionable changes. 
We present our work, \tool{}, an interactive, open-sourced web tool for visualizing these models' toxic predictions, while providing alternative suggestions for flagged toxic language. Our work also provides users with a new path of recourse when using these automated moderation tools. \tool{} highlights text responsible for classifying toxicity, and allows users to interactively substitute potentially toxic phrases with neutral alternatives. 
We examined the effect of \tool{} via two large-scale user evaluations, and found that \tool{} was highly effective at helping users reduce toxicity as detected through the model. Users also gained a stronger understanding of the underlying toxicity criterion used by black-box models, enabling transparency and recourse. 
In addition, we found that when users focus on optimizing language for these models instead of their own judgement (which is the implied incentive and goal of deploying automated models), these models cease to be effective classifiers of toxicity compared to human annotations. This opens a discussion for how toxicity detection models work and should work, and their effect on the future of online discourse. 
\end{abstract}



\begin{teaserfigure}
    \centering
    \includegraphics[width=0.8\textwidth]{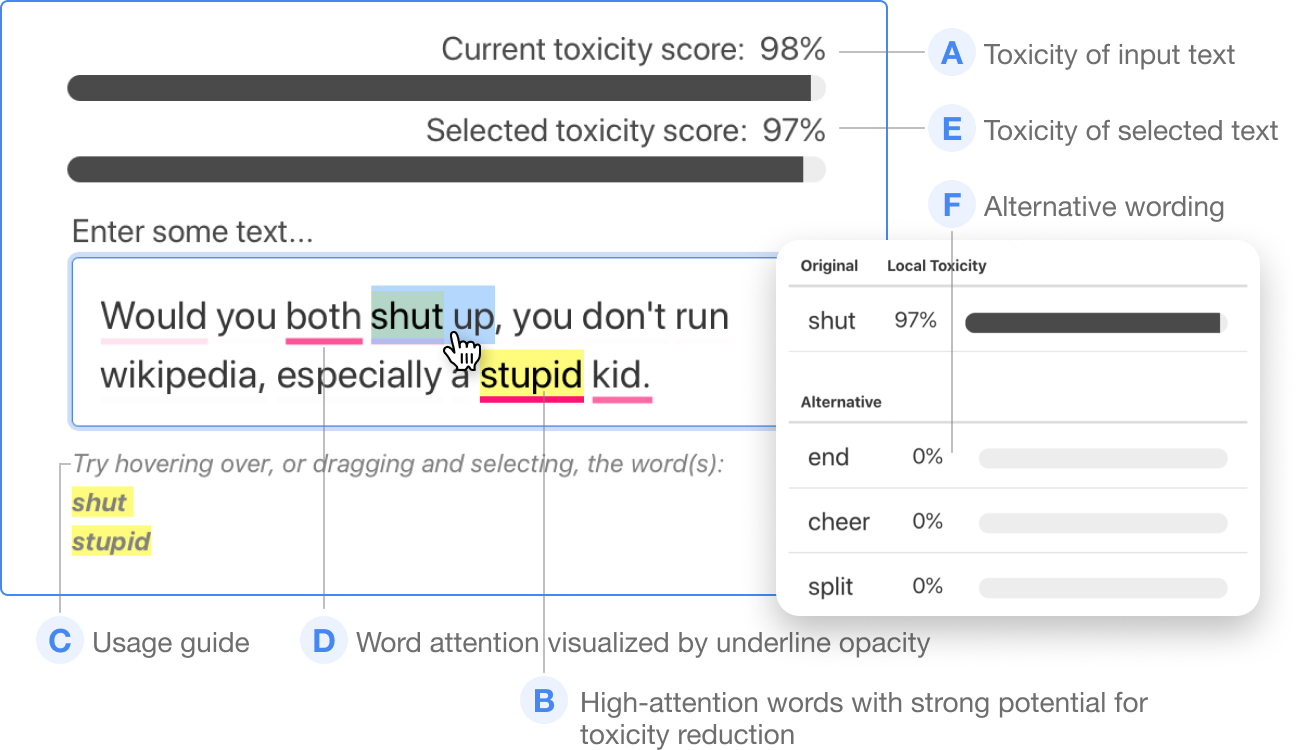}
    \caption{The \tool{} user interface. 
    \textbf{A.} Toxicity score of overall input text shows edits' effect on toxicity in real time.
    \textbf{B.} 
    Words whose possible alternatives have strong potential for toxicity reduction are highlighted in yellow.
    \textbf{C.} Usage guide for \tool{}'s capabilities. 
    \textbf{D.} Underline opacity visualizes model's attention on words,
    including those without alternatives, to inform users about which words contribute important context 
    (e.g., ``kid'' is underlined, because 
    toxicity towards a kid contributes to the toxic context.) 
    \textbf{E.} Showing the toxicity score of selected text  allows users to localize the sources of toxicity and search for the regions most important to edit.
    \textbf{F.} Hovering over highlighted toxic text displays alternative wording in a pop-up.
    }
    \label{fig:crownjewel}
\end{teaserfigure}


\maketitle

\textbf{\textcolor{orange}{ \small{WARNING: This paper contains some content which is offensive in nature.}}
}

\section{Introduction}


Toxicity online is widespread: a 2015 user survey on online social network platform Reddit found that 50\% of negative responses were attributed to hateful or offensive content~\cite{buni_chemaly_2016}; however, addressing toxicity through automated means is not trivial---as there must always be a choice of determining what language should be removed and what should not.
The same survey \textit{also} found that 35\% of complaints from extremely dissatisfied users were about heavy handed moderation and censorship. With the inherent trade-offs baked into content moderation, it is challenging to find a middle ground for this issue. Furthermore, the extreme scale of social media interactions \cite{brandwatch} exacerbates these challenges. These issues have lead to the development of automatic toxicity detection models such as the Google Perspective API \cite{perspectiveAPI}. 

Introducing automation, however, raises its own challenges. Machine learning models, responsible for detecting and moderating toxic language, can themselves be flawed \cite{sap-etal-2019-risk, hosseini2017deceiving}. When in the past users could rely on clear community standards from human moderation (or at the least an ability to communicate with a moderator), the adoption of fully automated systems make human-facilitated moderation much more difficult. Moderators who do not understand how automated tools work may not be able to contribute as much after these tools are adopted. As platforms rely more heavily on automated systems for moderation, users also receive less feedback
and might not be able to clearly connect their behaviors with community standards \cite{Jhaver_2019}. 
This fundamentally reduces the effectiveness of automated moderation, as users cannot learn what they did wrong especially if they are unfamiliar with the language or social norms of a platform. 


Without feedback or explanation, users of online forums that use toxicity detection systems based on black-box NLP models might question how their language is being examined. In such scenarios, there is no way to interpret \textit{why} the model considers language toxic. Without tools providing an avenue of recourse \cite{venkatasubramanian2020philosophical} designed for actual end users to audit what is being detected and make actionable changes to their language, people are disempowered to participate in discourse online. 
Furthermore, without the ability to detect when a model is falsely flagging language due to either linguistic limitations or social biases, the work of finding inaccuracies and correcting models is left entirely to the unrepresentative population of machine learning researchers and software engineers. 
Therefore, \textbf{providing end-users the ability to interactively audit the models that affect them will help democratize the improvement of these models.} 

Finally, interactive auditing opens an avenue for recourse. Given an explanation for how a model works, users can re-evaluate writing toxic text, increase awareness of potential limitations in toxicity detection models, and inform people who are unaware of certain toxic jargon. Black box models, however, are impossible to interrogate. These models provide end-users with little capacity to observe their underlying decision-making processes. Highlighting features that contribute to a model's output provides users with concrete evidence when pursuing recourse.

We address these challenges by developing an interactive tool called \textbf{\tool{}}, which allows for the interrogation of toxicity detection models through counterfactual alternative wording and attention visualization. \tool{}'s design does not require any expertise in machine learning from users, and 
enables them to visualize their language through the eyes of the algorithm. To this end, the primary focus of \tool{} is allowing users to visualize where and how a model detects toxicity within a specific piece of text, make actionable changes to their language to reduce toxicity \textit{as determined by the model}, and gain generalizable insights into how the model works to inform future language and spur potential changes if flaws are found. Furthermore, we study the effects of allowing users to experiment with \tool{}, analyzing how their language changes as they are more aware of the model. To sum up, our contributions are:

\begin{enumerate}
    \item \textbf{\tool{}}, 
     an interactive system allowing users to dynamically interrogate the classification of toxicity by visualizing its sources within a piece of text and examining alternative wordings. This provides users a method to understand and interact with toxicity detection models. 

    \item \textbf{Experimental Findings}  By using \tool{} as a means for users to understand models and optimize language more efficiently, we not only evaluate the effectiveness of \tool{}, but also study potential long term effects on discourse---caused by structural incentives from  automated toxicity filters. We find that as users learn to optimize their language (a necessity to avoid being censored as these models become more prevalent), human labeled toxicity can \textit{increase} compared to naïve editing. Thus, this work provides the first large scale quantitative evaluation of toxicity detection models and their potential effects on online discourse. 
    
    \item \textbf{Open Source Implementation} of \tool{} that enables broad access and future work. \tool{} provides a model agnostic framework for analysis of toxicity detection models by not only model developers, but domain users and human moderators. \tool{} can be used so that all relevant parties can become better aware of emerging issues in these models. We have uploaded our source code as supplementary material. We will immediately make the code public on GitHub upon publication of this work.

\end{enumerate}

\section{Related Work}
\subsection{Content Moderation}
There has been extensive research on human-managed online content moderation \cite{gillespie_2018, seering2019designing}, especially relating to the effect of transparency from the perspective of the end user \cite{jhaver2019did, Jhaver_2018_block} and the importance of explanations \cite{Jhaver:2019Transparency}. Various works have called for the development of tools that support, rather than supplant, the efforts of human moderators \cite{Jhaver:2019Transparency, seering_2020, seering2019designing}. Furthermore, several works have focused on studying the development of new automated moderation systems \cite{long_2017_bot, crossmod_2019}, and their effect on the dynamics of the platforms that utilize them  \cite{stuart_2010_vandal}. 
However, most of this research has focused on rule based systems, such as the Reddit Auto-Moderator \cite{Jhaver_2019}. In contrast, work that has used statistical machine learning approaches, possibly with the exception of \citet{crossmod_2019}, has often focused on accuracy and over transparency or moderator experience \cite{burnap2015cyber}. In this work we aim to study the effect of these systems from \emph{the perspective of the end user} as opposed to moderators managing the system. In particular, we look at the effects of deep neural network based moderation systems, which are notoriously difficult to interpret, in comparison to rule based systems. This also introduces issues regarding the perceived legitimacy of platforms, as neural models lack many of the core procedural values required for a platform to be seen as fair, such as due process, transparency, and openness \cite{suzor2018evaluating}. In some cases, linguistic variation also allows users to bypass automated content moderation systems \cite{10.1145/2818048.2819963}. Understanding what a model perceives as toxic is integral to evaluating its efficacy. Therefore, a key goal of this work \textbf{is to provide or outline implicit procedures used by automated toxicity detection models that can bring higher levels of transparency in these systems.} Recent work has shown that even limited transparency and explanations from automated systems can be as effective as explanations from human moderators, which ``suggest an opportunity for deploying automated tools at a higher rate for the purpose of providing explanations'' \cite{Jhaver:2019Transparency}. 
We also hope that, similar to \citet{matias_civil}, this platform can serve as a host for experiments in visualizing the predictions of different types of algorithms and the impact of these visualizations on user behaviour.

\subsection{Toxicity Reduction Interventions}
Many researchers have explored and built both social and technical approaches to identifying, reducing, and combating hateful content online \cite{strachan2014interventions,mathew2019thou, seering2019designing, jie_2019, chatzakou2017mean, Filippo2015MisogynisticLO}. 
One of the main solutions these research or platforms propose is to block, ban or suspend the message or the user account.
Although removing content or banning relevant users who are perceived as toxic by automated models may reduce their impact to some extent, it may also eliminate legitimate or important speech. 
A number of interventions have been developed in the CSCW/CHI community to combat hateful content and harassment \cite{mahar2018squadbox}. 
For instance, \citet{seering2019designing} designed a system 
using psychologically ``embedded'' CAPTCHAs containing stimuli intended to prime positive emotions and mindsets, influencing discussion positively in online forums, while \citet{taylor_empathdesign} explored cues that could encourage bystander interventions. 
\citet{zhuEstelleWiki} raises concerns about conflict between automated moderation systems and human guidelines, studying Wikipedia's current automated moderation system. 
\citet{crossmod_2019} introduced a sociotechnical moderation system for Reddit called Crossmod to help detect and signal comments that would be removed by moderators. 
In contrast, our proposed tool \tool{} aims at influencing discourse more directly at the end-user stage to promote fairness and interpretability. 
\citet{citron2011intermediaries} examined a number of efforts on hate speech and identified three ways of responding to hate speech: (1) removing hateful content, 
(2) directly rebutting hate speech, and
(3) educating and empowering community users. 
Our proposed tool aligns well with (3), as \tool{} enables users to see the toxicity levels of their content transparently, offers examples of instances when content is and is not toxic, and provides model visualization and reasoning. Furthermore, \tool{} helps users with alternative wording, helping them express similar ideas in non-toxic ways.

\subsection{Natural Language Models and Toxicity Detection}

Driving the development of many new toxicity detection methods are recent advances in the field of NLP. In particular, the introduction of massive pretrained Transformer \cite{vaswani2017attention} models such as BERT \cite{devlin2018bert} have accelerated the state of the art in many downstream tasks like toxic language classification. 
The learned representation from these models are often used as the input for a relatively simple model which can be trained on a specific task like toxicity classification. This process is very common as it hugely reduces the amount of training and data required to achieve good results. However, this does mean that any potential issues present in BERT (or other pretrained Transformer models) are inherited by the fine-tuned model. Issues include bias  \cite{bolukbasi2016man, sap-etal-2019-risk, manzini-etal-2019-black}, as well as a general lack of semantic understanding \cite{ettinger2019bert}. Such issues are particularly important when considering notions of toxicity, which can take many forms, some of which are linguistically pleasant but semantically abhorrent. In fact, work has shown that the Perspective API is susceptible to the same adversarial attacks that fool other NLP models  \cite{hosseini2017deceiving}. To address this, there has been some progress on building frameworks to automatically mitigate bias in text directly \cite{pryzant2019automatically}; however many problems in this space remain open. Despite these issues, deep neural network based models outperform more traditional rule based models in detecting biased language \cite{hube_2019}, and are increasingly being deployed and thus carrying these flaws into socially-important real-world situations.  
\subsection{Visual Language Interpretability Systems}
In order to express information about the model to non-technical end users, our tool fits generally within the tradition of visual analytics for deep learning explainability \cite{hohman2018visual}. From a visual analytics perspective, various interactive tools have been built for understanding the internal mechanisms of general purpose natural language processing systems. However, (a) there has been little work on understanding the function and impact of toxicity detection systems specifically and (b) these tools are aimed mainly towards developers. Some tools such as SANVis \cite{park2019sanvis} and exBert  \cite{hoover2019exbert} allow for interactive exploration of the attention mechanisms in Transformer models. The attention scores associated with each word allow connections between words and emphasis of certain words in context to be highlighted. These approaches take a generally static view of the data, where the tool is viewed as a method to explore existing data.

However, dynamic visualizations that allow for user experimentation can assist in the understanding of a machine learning model \cite{lai2020chicago}. Some tools take a more active approach to explain models through counterfactuals \cite{Wexler_2019_whatif}. Others, like Errudite \cite{wu-etal-2019-errudite}, allow users to test their own hypotheses with respect to the true error distribution on the entire dataset. Some techniques attempt to adversarially perturb language input to a model \cite{zhang2019adversarial, slack2019fool} in order to change its classification; other methods use a human-in-the-loop design  \cite{br2019visual}. In this work, we synthesize these two paradigms by passively visualizing model attention, while also enabling interactivity with AI-guided and human-driven counterfactuals. Importantly, we design our tool in the context of automated moderation systems, prioritizing usability from the perspective of non-specialists.

Finally, tools like The Perspective API also offer limited interpretability, allowing platforms to embed text editing areas with a small widget that notifies users with a binary output (toxic/non-toxic) when their input exceeds a toxicity threshold. It also updates as users edit, and allows easy feedback for users to notify that they think the model was incorrect. This provides some of the benefits of \tool{}, however it lacks more extensive visual information to help lead users to specific problems in their text, which maintains the black box nature of the model.

\section{Design of \tool{}}

In this section, we motivate the design of the \tool{} tool through a formative user survey, and formalize a series of design goals from the concerns raised in our user survey.

\subsection{Formative Survey for Understanding Automated Content Moderation}\label{section:user_needs}

To understand difficulties relating to toxicity moderation, and to outline user needs (especially those related to recourse), we surveyed 100 Amazon Mechanical Turk Workers with social media accounts in the United States about their experience surrounding online toxicity and moderation.\footnote{The survey took an average of 6 minutes to complete with compensation of \$0.80, above the US Federal minimum wage. Workers were selected from the Amazon Mechanical Turk pool with filters to ensure they were within the United States, and held Reddit and Twitter accounts to ensure a certain familiarity with online discourse. Respondents had an average age of 34 (standard deviation of 8 years), identified as 66\% male, 33\% female, and 1\% nonbinary, and 75\% White, 9\%
Asian, 8\% black, 2\% Latino, 1\% Native American, and 5\% other/unspecified.} 

We asked how often users noticed toxic language on the internet and found that 62\% responded noticing toxicity either `often' or `always' compared to only 6\% responding with either `rarely' or `never.'
Our survey also highlights the clear effect of racism as an undeniable component, with 16\% of non-white users reporting `always' noticing toxicity compared to only 4\% of white users (and only 2\% for white males). Toxicity is not only noticed but caustic: we asked about the effect toxicity had on users lives both online and offline, and found that overall 43\% responded that toxic language online had a `somewhat negative' or `very negative' effect on their lives offline. This effect was also influenced by gender, with only 20\% of males reporting negative effects while 52\% of non-males reporting negative effects offline. We find that not only is toxic language pervasive but it disproportionately and intersectionally harms already underprivileged groups.  

Given the groups' overall clear negative experience of toxicity, we also wanted to understand how they felt about the trade-off between free speech on online spaces and toxic language. In an open ended question, 36 responses were highly supportive of strong moderation to reduce toxicity, with responses along the lines of \textit{``I think the tradeoff is well worth it.  I am so tired of hearing foul language all the time.''} At the same time 31 were very skeptical of any moderation and highly supportive of user freedom of speech, responding \textit{``I think free speech is important. So people can say what they want even if it is toxic.''} These divided responses highlight the inherent difficulty of balancing the standards of content moderation and freedom of speech. 

Many neutral responses were additionally skeptical of automated systems in particular. For example, one participant noted that \textit{``in many platforms moderation is biased. The automatic tools are not sufficient to remove toxic material from the site. These tools end up removing quality content instead of toxic ones.''} A common complaint from these responses claimed that addressing the issue of toxicity through simple models would exacerbate existing tensions between moderators and users. To understand the specific problems users had with these automated systems we then asked about feedback. Among participants who have had a post removed either by a moderator or an automated system, we found that when removed by a human, 52\% reported receiving feedback `often' or `always', while only 36\% reported receiving feedback from automated systems.

Finally, to understand which areas users felt need the most improvement, we asked participants how they would improve existing systems for online content moderation, and what features they would want in a tool. Concretely, users noted wanting familiarity and similarity to existing tools for modifying language in spellcheck and auto-complete interfaces \textbf{``\emph{like auto correct but for language vs. typos}''} that could simply \textbf{``\emph{suggest other words}''}. 

This aligned with a desire that any tool should be \textit{``something friendly and approachable that isn't too intrusive or annoying.''} Many of the responses were outright hostile to the concept of a tool for reducing toxicity for fear of censorship, meaning that any such tool would be most effective the less visible it is---similar to the notion of nudging in behavioural economics~\cite{thaler2009nudge}.

Stemming from the same well justified fear, users emphasized that such a tool requires \textit{``an appeals process, information about why a post is removed, a rejection of a post with advice for fixing it before posting something''}; and \textit{``the ability to give feedback on the tool since it will almost certainly have failure scenarios. The tool should also be able to work in real time, and have an excellent understanding of English.''} From our survey, we found that users indirectly recognize that current machine learning systems do not have a perfect understanding of natural language. Thus, a more concrete way to provide feedback is very important for such a system. Finally one user noted that they would like to see the sources for a model, \textit{``guidelines.. lots of written comments of what is deemed toxic''}, which helps justify the need for these models and datasets to be open source. 

\subsection{Design Goals}

We synthesized the information from the study to identify the main design goals for \tool{}.
\begin{itemize}[topsep=3mm, itemsep=2mm, parsep=0mm, leftmargin=10mm]
 \item[\textbf{G1}] 
    \textbf{Interpretation.}
    Users often feel as if they are unsure of the specific guidelines and requirements they are asked to maintain, and may not know what about their language may cause a post to be removed. Therefore it is important to provide explanations to users that are useful not only for a specific comment but more generally for how the classifiers work. Ease of interpretability will enable users to build appropriate mental models for planning how they use language in the future. These interpretations can provide the bedrock of any potential action a user may want to make either on or off of the platform. 
\item[\textbf{G2}] 
    \textbf{User Driven.}
    In order to make sure users do not feel overly censored, we ensure that no decision about editing text is to be made without the explicit choice of the user, and that a wide variety of options be presented to maximize the capability of the user to say what they mean. This differentiates \tool{} from fully automatic, end-to-end approaches, which have been presented for reducing bias or toxicity \cite{pryzant2019automatically}. A trade-off with this design principle concerns users who are determined to use toxic language. However, such users would not use a tool like this in the first place if it prevented their ultimate desired language. In order to understand how real users might use similar tools, and to study how even toxic users interact with moderation models, we prioritize a user-centered design.
\item[\textbf{G3}] 
    \textbf{Minimalism.}
    In order to ensure accessibility for end users who are not familiar with complex data visualization paradigms, and to make sure that the tool is not overbearing or irritating, we aim to build a tool that minimizes extraneous views. This also has a trade-off of precluding more advanced or comprehensive user interfaces; for the purposes of this study, however, providing greater accessibility for users (for instance users whose first language is not English) is an important consideration and thus is prioritised.  
 \item[\textbf{G4}] 
    \textbf{Easy Feedback.}
    Anticipating that any model for detecting toxicity will be flawed, it is valuable for users to be able to highlight erroneous classification easily to ensure they feel they are being heard, and to improve the underlying model when possible. 
\item[\textbf{G5}]
    \textbf{Accessibility.}
    To develop a tool that is accessible for users without specialized computational resources,
    we deploy our tool using lightweight modern web technologies, and place emphasis on ensuring our system runs efficiently for low-resourced users. We also open-source our code to support reproducible research.
\end{itemize}

\subsection{Ethical Considerations}\label{sec:ethics}

Outside of the primary design goals of \tool{}, there are special considerations we need to give to potential ethical issues\footnote{These considerations are in addition to standard practice considerations with anonymous data collection and annotations. This research study has been approved by the Institutional Review Board (IRB) at the researchers' institution.}.
\tool{} enables users to edit text so that it is no longer be detected as toxic by a classification model. However it is highly possible that the resulting text can in reality remain toxic. Aspects of this work are similar in functionality to work done to generate adversarial examples for text classification \cite{Li_2019,br2019visual}. Bad actors could potentially use \tool{} to pass truly toxic language past existing filters.

To avoid these scenarios, we have included controls within \tool{} that prevent it being used in such a scenario where it detects explicit hate speech or uneditable/irredeemable toxicity. Many bad actors already have the ability to bypass models through the method of trial and error. The end result of proliferation of a tool like \tool{} is not to improve the ability for bad actors to `game the algorithm,' as it is already being gamed \cite{10.1145/2818048.2819963}. For example, the \#thyghgapp vs \#thighgap phenomena, outlined by \citet{10.1145/2818048.2819963}, highlights how bad actors already change syntax to avoid detection from automated models. Rather, \tool{} aims to provide a novel service to users who are already acting in good faith to provide transparency about automatic moderation methods. This set of issues is not unique to \tool{}, instead these issues implicate the toxicity detection models themselves, and the platforms deploying them. Users cannot demand or instigate change if they do not understand these issues, which further motivates the development of \tool{} despite its initial risks.

Finally, a primary contribution of this work is not just the development of \tool{} for its own sake, but also as a means of understanding how toxicity detection models affect people's choice of language when compared to human labeled toxicity. We suspect that optimizing language for automated models are an inevitable consequence of their increased deployment (as language approved by such models will become the only kind of language made visible). Because of these consequences, and in the effort to prevent any malicious use, our released open source contribution will include an informal consent form that discusses appropriate ethical issues, regulations, and best practices. 

\section{\tool{}}

\tool{} (\autoref{fig:crownjewel}) is an online interactive tool with the primary focus of allowing users to visualize toxicity within a text, make changes to reduce toxicity, and gain generalizable insights into how toxicity classifier models work.

 At the same time, it is important to note what \tool{} is \textit{not}. \tool{} is \textit{not} designed to necessarily change, explain, or interpret anything regarding ``real'' toxicity as far as it even can be directly or objectively identified. Rather \tool{} allows users' access to, interpretation of, and interaction with, \textit{toxicity detection models}. As we have established,
 the current state of the art of NLP can be linguistically naïve~\cite{ettinger2019bert}, thus we expect that notions of ``true'' toxicity and detected toxicity will diverge, especially in the scenario when users try to edit toxic language to comply with these models. Furthermore it is also important to note that the contribution of \tool{} is not novel NLP or visualization methods (in fact the \tool{} architecture is model agnostic), but rather in the synthesis of existing tools in addressing specific user issues and then studying the effect that the systems underlying these issues have on online discourse and user experience. 

\subsection{Visualizing Comment Toxicity}
The primary information \tool{} expresses is the toxicity classification score of user input. Users can input text and view the toxicity of the overall sentence, along with which words contribute most to the output score. 
\tool{} displays a score between 0 and 100, which represents the probability that the model assigns to the input for whether or not it is toxic. This is shown at the very top of the tool as a bar (\autoref{fig:crownjewel}A). The minimal bar design, in monochrome and removed somewhat from the text, is noticeable enough to provide informational feedback, while subdued enough to not distract from the text itself. 
The toxicity bar dynamically updates as the user edits the text. This allows users to experiment with their own or suggested edits and get real time feedback for any counterfactual scenarios. This enables both effective exploration to choose the best possible wording and easy iteration to test hypothesis for how the model works. The inclusion of this metric visualization may incentivize a local optimization or `hill-climbing' approach. However this approach better aligns with the user desire for limited intervention, as a user may be more likely to make an edit if it is a small change to their text rather than complete rewrite. Furthermore the reactive visual provides immediate feedback, facilitating experimentation which has shown to be effective in building understanding~\cite{lai2020chicago}.

\begin{table}
    \centering
    \resizebox{0.75\textwidth}{!}{
    \begin{tabular}{l r r}
        \toprule
        Explanation Method & Human Annotation Overlap & Average Compute Time\\ 
        \midrule
        Integrated Gradient Based & $.86 \pm .06$ & 1880 ms \\
        Attention Based & $.87 \pm .07$ & 99 ms\\ 
        \bottomrule
    \end{tabular}}
    \caption{Comparing gradient and attention based methods for flagging toxic words in \tool{}. Evaluations were run a computer with an Intel i7 2600K and a single NVIDIA GTX 1070.
    }
    \label{tab:user_attn}
\end{table}

\subsection{Explaining Toxicity Classification}
\label{toxicity-attn-just}
A key component of \tool{} involves identifying the tokens in a text that are indicative of toxicity, and attributing importance to these tokens. There are two widely-used automated techniques that perform attribution: gradient based explanation and attention based explanation \cite{vaswani2017attention, sundararajan2017axiomatic}. Both these techniques offer a numeric value for each word in an input sentence, where the magnitude of the numeric value corresponds to relative importance in a model's prediction. However, the capacity for these methods to explain model predictions may differ across tasks \cite{wiegreffe2019attention}. In this subsection, we compare gradient and attention based explanations to human annotations and to each-other, selecting an appropriate technique for \tool{}'s backend. 

Although there is much debate as to whether attention is a good proxy for explanation~\cite{wiegreffe2019attention}, when interpreted carefully, attention can be a \emph{rough and weak proxy} for explanation \cite{Clark_2019, serrano-smith-2019-attention}. As a precaution, we compared our attention based metric (denoted as $attn$) to typical gradient/saliency based techniques (denoted as $grad$). For $attn$, we computed the  average attention score over last layer heads in our Transformer model (before the linear classification layers) using the end/CLS token on the input text. On the other hand, $grad$ was calculated using an integrated gradient approach documented by \citet{sundararajan2017axiomatic}, using the standard gradient operation on the input to the model to evaluate importance. For evaluation purposes, we collected two different metrics: speed of inference, and set overlap ($\operatorname{overlap}(X, Y)=\frac{|X \cap Y|}{\min (|X|,|Y|)}$) for flagged words in sentences. In the set overlap metric, $X$ and $Y$ are sets of flagged words from a given sentence.

To effectively compare both our techniques, we aligned differing output ranges for $grad$ and $attn$, since $attn$ is bounded between $[0, 1]$ and $grad$ between $[0, \infty]$. We manually tuned cutoffs for attention (i.e. $attn$ $> .2$) and found cutoffs for gradient based approaches by collecting the distribution of $attn$ scores, $p(x)$, and $grad$ scores, $q(x)$, over a random 5\% subset of our training dataset (7979 instances). Then we took the percentile value of our $.2$ $attn$ cutoff at $P^{-1}(.2) = .9$, and we identified the corresponding gradient cutoff using the percentile value from the attention distributions, $q^{-1}(P^{-1}(.2)) = .02$. Finally, we selected words from a smaller subset (50 instances) to compare attention and gradient based flagging to human annotations.

We conducted two analyses to compare gradient and attention based methods for explainability:
\begin{enumerate}
    \item Analyze word overlap and inference speed on a random subset of our training data (7979 instances), comparing $grad$ and $attn$.
    \item Analyze word overlap on a smaller random subset of our training data (50 instances), comparing $grad$, $attn$, and human annotations. The guidelines for this task required annotators to flag all words contributing to toxicity either implicitly or explicitly.
\end{enumerate}

\textbf{For analysis 1,} we found that the average $overlap(grad, attn) = .82 \pm .02$ at a 95\% confidence interval, for 7979 instances. For inference times, saliency methods required significantly more time due to the added back-propagation step. We recorded flagging speed per batch, with $grad$ at 1.88 s $\pm$ 9.92 ms (mean $\pm$ std. dev. of 7 runs, 1 loop each), and $attn$ at 98.8 ms $\pm$ 2.46 ms per loop (mean $\pm$ std. dev. of 7 runs, 10 loops each). \textbf{For analysis 2,} the authors who are familiar with the task setup annotated 50 random examples manually, highlighting tokens they considered toxic. We recorded average $overlap(grad, attn) = .79 \pm .12$, $overlap(grad, human) = .86 \pm .06$, and $overlap(attn, human) = .87 \pm .07$ -- all values are at 95\% confidence intervals. Regardless of flagging technique, we find that each method highlights core toxic elements in text that align with human annotations. 

In order to achieve real-time explanations with minimal latency, gradient explanations are prohibitively slow ($1.88s$ vs $98.8ms$). In our token flagging task, both attention and gradient techniques perform similarly, and have reasonable overlap with human annotations. Therefore, we use the attention to explain which words are highly associated with our model's predictions. We importantly understand that attention, in some contexts, may not explain model prediction. However, in our specific scenario (flagging toxic phrases), attention is both \textbf{significantly faster and flags similar tokens} when compared with human annotation (overlap at $.82$ and $.87$ for both task 1 and 2, respectively). Because our selected techniques perform similarly, and $attn$ provides improved inference speeds,  we utilize $attn$ for this work.

\subsection{Visualizing Model Attention}

Various visualization concepts have been used to show the relative importance of words and visualize attention, such as \hl{highlighting} and \textcolor{gray}{opacity} \cite{attn_vis_notebook}. However, we utilized an \setulcolor{red}\ul{underline} on every word, where the opacity of each underline would be controlled by the magnitude of attention placed on each word (\autoref{fig:crownjewel}D). This method was chosen as it helped with legibility of the text, which is vital for users understanding differences in textual classifications. Like the classification bar, its design is purposefully simple, mirroring the text interaction techniques of underlining to note where editing is required---something end users are familiar with from common text editing software.
This accessibility allows \tool{} to effectively communicate the complexities of toxicity detection models using a visual language users are already fluent in. 

\begin{figure}
    \centering
    \includegraphics[width=\textwidth]{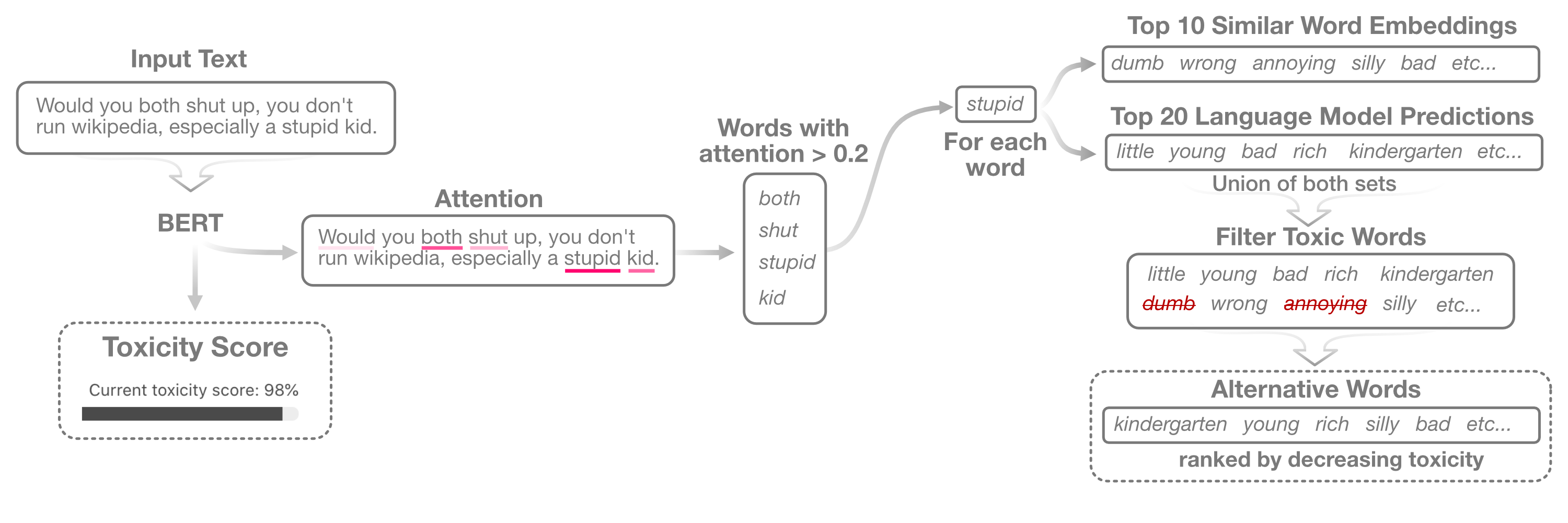}
    \caption{Process for generating alternative words}
    \label{fig:end_to_end}
\end{figure}

\subsection{Alternative Wording}
Beyond passive visualizations, \tool{} suggests concrete, actionable edits to text, assisting users in lowering toxicity through alternative wording. The alternative wording feature provides users with options to swap or delete words in a sentence that are responsible for high toxicity scores. A word in the input is highlighted (\autoref{fig:crownjewel}B) to draw particular attention to it when it meets the criteria of: 
\begin{enumerate}
    \item An attention score greater that 0.2
    \item \tool{} can find alternative words with individual toxicity less than 0.4
    \item The alternatives have a positive impact on the overall input toxicity.
\end{enumerate}
These thresholds were determined during the analysis to best match the attention highlights to human annotation. The requirements of alternatives having a positive impact both globally (for the whole text) and locally (for the replacement word on its own to be benign) provides a safeguard against malicious use.

When the user hovers over any of these words, suggested substitutions are shown and ranked in a popup (\autoref{fig:crownjewel}F). Selecting one of these alternatives replaces the word and the new toxicity score is updated. This mode of interaction is also easy and intuitive for users due to its similarity to familiar spellcheck or thesaurus tools (motivated by our survey), requires little retyping of edits, and gives options if users cannot immediately think of an alternative word. Furthermore, it displays a range of options which gives the end user agency in maintaining the original meaning as closely as possible. Finally, beyond the act of making the sentence less toxic, the technique allows users to learn which words tend to be highlighted, and what common synonyms the algorithm tends to suggest. This allows people to learn about the model and use this knowledge while writing future comments. 

\autoref{fig:end_to_end} illustrates how the set of alternative words for a given toxic input word is calculated. Given a word for which we are calculating potential alternatives we first find its nearest neighbors in a Glove \cite{jeffreypennington2014glove} word embedding space \cite{10.5555/2999792.2999959}. We limit nearest neighbor search to 10, balancing the amount of user choice in word options while reducing the cognitive load of choosing among too many \cite{tufte_2018}. These words will match the original word closely in meaning, and provide a solid base of options for users to reword from. Furthermore these vectors are not dependent on a manually curated list of synonyms and extends to find similar but non-synonym words. Next we use a BERT language model to find other words that may fit within the context of the word to be replaced. We do this by feeding the original input into the language model with the selected word masked, causing the language model to output a probability distribution of likely words that fit within that context. From these words, we select the 20 most likely. Then, we take the union of these two sets and filter out any words with individual toxicity greater than 0.4, along with words that do not have a positive impact on the overall input toxicity. 

Using current state of the art NLP models does not ensure that every option will be a good replacement. To this end, \tool{} highlights several alternatives so the user will likely find at least one good replacement that they can select. Our controls ensure that no replacement makes the result worse. This gives the user the most amount of control over the process while still leveraging all of the potential options and power of modern NLP systems. 


\subsection{Multiple Alternatives}
In addition to replacing single words, sometimes toxic words come in groups or phrases where the toxicity is not individually attributable to any one of the words, this may require a different editing paradigm for the end user.
To support this, \tool{} allows a user to select a contiguous text phrase, and alternatives are generated for the n-gram of toxic words contained within that phrase. Sets of words are chosen from the same universe of words as for single word replacements through word embedding similarity. Furthermore, the language model masks the entire set of words to be replaced and gets the most likely tuples from the resulting joint distribution. This allows \tool{} to encode the linguistic coherence of not just each word in context but the whole set of words within context. Finally sets of words are ranked by the resultant toxicity of the edit on the selection as before. 

\subsection{User Feedback}
Knowing that the classification model is expected to make mistakes, and that a goal is to provide user recourse for handling those mistakes, we have also included an integrated feedback form within the tool. If a user feels the model has made a mistake, there is an included text box below the main input space for comments to submit to the developers. In a deployed system, this would forward complaints on to the relevant platform. This can be used as a means for re-training and improving the model as well as providing a direct way for users to pressure platforms when models exhibits bias or other issues. By logging inaccuracies highlighted using \tool{}, end users can concretely identify when models utilize tokens that should not be attributed with a toxic prediction. Furthermore, visual feedback from \tool{} provides developers and researchers with identifiable sources of errors in their models.

\subsection{System Implementation}
While \tool{} as a tool is built to be model agnostic (as long as a model uses attention or similar method), for our evaluation we needed to include a backend implementation of current state of the art toxicity detection models as a useful proxy for deployed systems. 

\subsubsection{Dataset}
The dataset we used for training the backend model for \tool{} was sourced by the Kaggle competition run by Google's Jigsaw \cite{jigsaw}, which is based on the dataset used by Google's Perspective API \cite{perspectiveAPI}. The Perspective API is used as one of the most commonly used and openly available content moderation tools. Therefore, its underlying dataset was a suitable proxy for \tool{}'s goal. By benchmarking against this dataset, we can compare our performance directly against that of the Perspective API, and thus be well justified in the representativeness of our model. 

We also chose to use this dataset because it was pre-cleaned, openly available, and contains a wide variety of baseline models through the Kaggle competition. The dataset consists of a set of 312735 comments from Wikipedia’s talk page edits, along with multiple labels that characterise the form of toxicity (toxic, severe toxic, obscene, threat, insult, and identity hate). We chose to only use the toxic label in the dataset for modeling purposes, as the other labels were subsets of toxicity and we wanted a sharper focus. 

A noteworthy limitation of this dataset is its focus on explicit hate speech, or speech that directly insults through the use of particular keywords and phrases (like ``shut up'' and ``stupid,'' as seen in \autoref{fig:crownjewel}). Implicit hate speech, however, tends to focus on stereotypes, avoiding explicit phrases (e.g., ``you're smart for a girl'' containing no individually toxic components yet expressing a misogynist meaning). Future work on collecting implicit hate speech is needed to help extend \tool{} and other toxicity detection systems to support such examples. 

\subsubsection{Model Architecture}
To detect toxicity in text, we fine-tuned a state-of-the-art Transformer based model (BERT) that performs reasonably well across various language modeling tasks.  Transformer based models rely extensively on self-attention to predict text \cite{vaswani2017attention}. 
Concretely, self-attention allows a model to detect toxicity based on the context of a word. Transformer models work by applying self-attention mechanisms to the input several times, over several layers. A final output is selected by propagating attention across these layers.  Although there are a wide range of possible models, such as those proposed in the original 2017 Jigsaw Kaggle competition \cite{jigsaw}, we decided to utilize Transformer models due to their prevalence in most modern NLP tasks, as \tool{} aims to be generally useful for current models.

\autoref{tab:baselineperf} summarizes performances across several baseline classifiers using the ROC-AUC metric~\cite{ROC_2001}, which is a standard metric in machine learning classification problems and the one reported in the Kaggle leaderboard. The BERT model we used performs on par with the current state of the art Kaggle leader-board by utilizing a single model as opposed to an extremely opaque ensemble of models. Furthermore, it remains representative of the current trends in NLP.

\begin{table}
\centering
\small 
\begin{tabular}{@{}lr@{}}  
\toprule
Model &  ROC-AUC \\ 
\midrule
Logistic Regression & 0.963   \\
Naïve-Bayes SVM & 0.972 \\
\begin{tabular}[t]{@{}l@{}}LSTM\end{tabular} & 0.977\\
\textbf{Fine Tuned BERT (\tool{})} & \textbf{0.982}\\
Large Ensemble (Kaggle Leader) & 0.989\\
\bottomrule
\end{tabular}
\caption{Kaggle Reported Toxicity Detection Performance. Highlighting the fine tuned BERT model used in this work.}
\label{tab:baselineperf}

\end{table}

\subsubsection{Software}
All deep learning based models used in our system were implemented in PyTorch \cite{NEURIPS2019_9015}, a library for building deep neural networks. We also utilized the HuggingFace package for pretrained Transformer models. Although we use Transformers across \tool{}, we built our frontend to be model agnostic, and so the backend model can easily be swapped out without major code changes (provided the replacement model supports generating attention-like explanations for its predictions). Our frontend  was written using Svelte for compartmentalizing our code, and D3.js for miscellaneous visual elements. Because \tool{}'s predictions occur on the backend, our client itself can run on systems with reduced computational power. \tool{} does not present any novel NLP architectures, techniques or methods. Instead, it is built upon the current state of the art, with consideration for real time performance. A contribution of this work is the usable tool itself as an implementation designed for this specific use case, and the insights gained from being able to study users interacting with the tool.

\section{Evaluation}
\begin{figure}
    \centering
    \includegraphics[width=\textwidth]{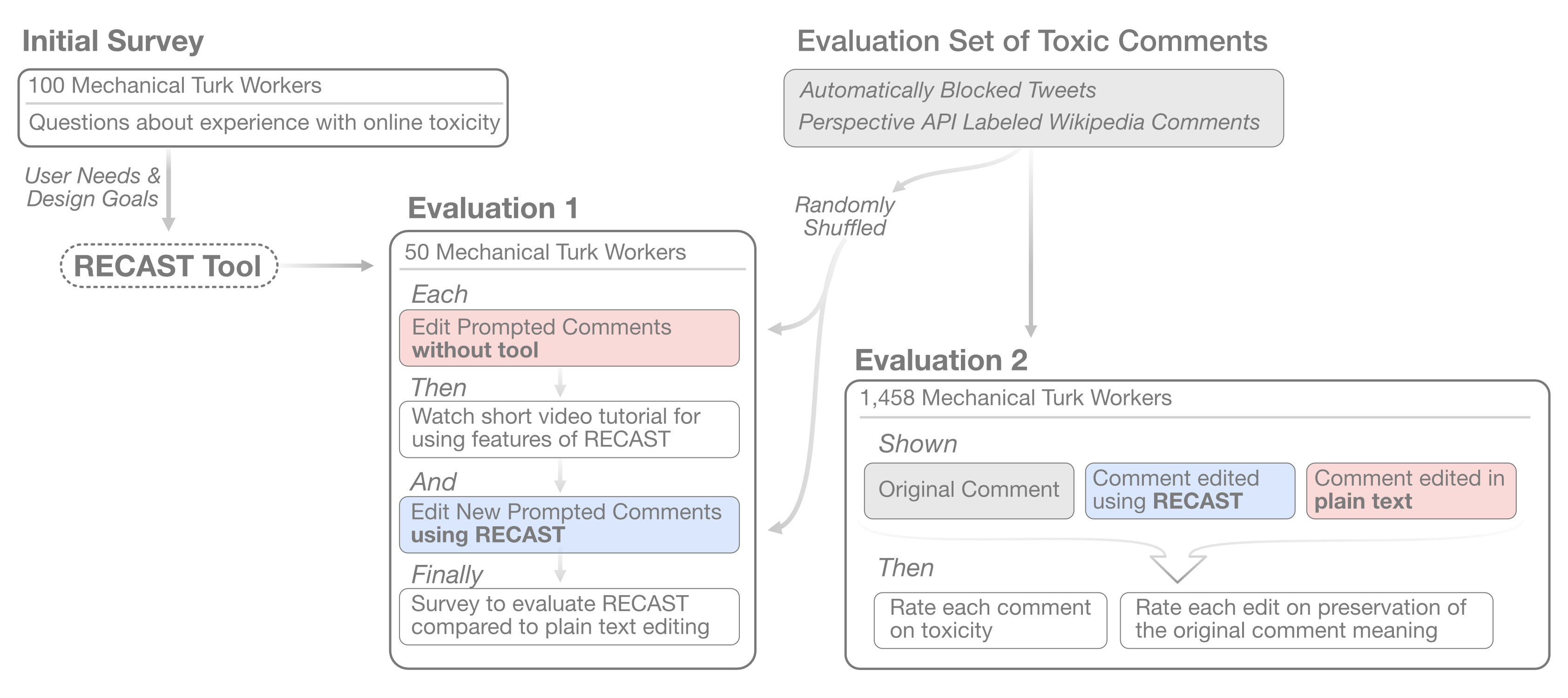}
    \caption{Multiple Evaluation Procedure}
    \label{fig:study_fig}
\end{figure}

We conducted two coordinated evaluations of \tool{}, as outlined in \autoref{fig:study_fig}, to study how well it addressed our design goals as well as to study the effect that user interpretability of toxicity detection models might have on online discourse. 
\subsection{Evaluation 1: Editing Toxic Comments with \tool{}}

\subsubsection{Methodology}
In order to evaluate how users would use a tool like \tool{}, we considered the task of editing toxic comments on social media. This scenario is representative of a situation a user would be presented with when using a tool like \tool{}---users would be interested in interacting with the model and potentially making changes to their original comment only after their commented is detected as toxic. Since one of our insights from the initial study was that users would prefer a tool to be lightweight, we imagine that \tool{} would only be deployed in these kinds of editing situations, where a comment is already recognized as toxic instead of being used every time a user writes from scratch. 
As a result, we present annotators with potentially toxic comments compared to asking annotators to come up with some on their own. 
While there may be some difference between users editing provided comments instead of editing their own comments, having a common set of comments for users to edit and thus comparable resulting outputs provides a larger benefit in terms of reproducible analysis. 

\subsubsection{Sample Pool}
For this evaluation we conducted a within-subject study which compared editing comments using \tool{} to normal editing without any help from \tool{}. 
We recruited 50 users (with a mean age of 38, consisting of 32 self-identified males and 18 self-identified females, all within the United States), from Amazon Mechanical Turk, an online microtasking platform. Users were paid \$1.80 per task (matching the United States federal minimum wage).

\subsubsection{Task Description}
The task users were given was to edit a set of provided comments scraped from two sources. The Perspective Kaggle dataset \cite{jigsaw}, consisting of Wikipedia comments manually labeled as toxic, provided a strong source of comments for model training. To supplement our study with comments outside of the Kaggle dataset, we took replies on Twitter scraped on May 9, 2020, found by looking at top trending hashtags in the US. We collected responses that were hidden below all other replies, behind the following filter warning: ``Show additional replies, including those that may contain offensive content.''\footnote{Specific comment text for each of these cases used in the user study can be found in the appendix.}
This mixture of sampled comments allowed us to examine both examples within and outside of the training dataset. In the case of Twitter, we provided users with contextual thread information to help them decide how to edit the comments. For a given editing task, users were provided a description of the context of the original comment, as well as preceding comments in the cases where they were available. 

Users were initially prompted to edit 4 comments using a standard text editor, then shown a video describing the features of \tool{}. Next, users were asked to edit a second set of four comments using \tool{}. In order to compare the resulting texts, we randomized the set of comments provided to the \tool{} enabled and \tool{} disabled groups. Finally users were asked to rate on a five-point Likert scale the degree to which they thought the addition of the tool did or did not help them reduce comment toxicity and understand the mechanics of the toxicity detection model. \autoref{tab:user_evals} shows that strong majorities found the tool to be easy to use and helpful, and provided a good understanding of the model's heuristics.

\begin{table}
    \centering
    \resizebox{0.95\textwidth}{!}{
    \begin{tabular}{l r r}
        \toprule
        Survey Question &  Proportion Agrees & 95\% CI\\ 
        \midrule
        The tool is easy to use. & 78\% & $\pm$12\% \\
        The tool is helpful in reducing the toxicity of a comment.& 70\% & $\pm$13\%\\ 
        The tool is helpful in understanding the criterion of labeling comments as toxic. & 80\% & $\pm$11\%\\
        \bottomrule
    \end{tabular}}
    \caption{User evaluations of \tool{}. Both ``Agree'' and ``Strongly Agree'' are included as agreement.}
    \label{tab:user_evals}
\end{table}

\subsubsection{Open Ended Responses}\label{sec:open_ended_Responsed}
To further validate these results and account for positivity bias in the responses, and to analyze the effectiveness of \tool{} in user understanding, we asked open-ended questions about what they learned about the model to gauge the generalizability of the patterns they learned through the study. We asked: \textit{``After using the tool, how would you characterize by what criterion language gets labeled as toxic versus benign?''}.

Many users noticed the tendency of the model to focus more on specific keywords than overall sentiment, as two users noted:
\begin{quote}
    \textit{``For the most part they get labeled by individual words with a negative connotation.''} 
\end{quote}
\begin{quote}
    \textit{``I think that it tends to pick keywords that can be considered highly offensive.''}
\end{quote}
Some users pointed out how keywords extended beyond just directly toxic words but common co-occurrences as well: 
\begin{quote}
    \textit{``I think that language that is obviously offensive (slurs, etc.) is labeled as toxic, as well as words that, with a high frequency, occur often close to other offensive words to make up larger phrases.''}
\end{quote}
However some users noticed the cultural influence and flaws in which words were highlighted: 
\begin{quote}
    \textit{``I think slang words and curse words are flagged more than negative opinions.''}
\end{quote}
\begin{quote}
    \textit{``I think that, especially slang, gets misconstrued within the tool and they falsely label it as toxic, when in reality it's not.''}
\end{quote}
One user also noticed the flaws of the underlying model by experimenting themselves with the tool, as they noticed differences between toxic language that the model highlights, and toxic meaning which it does not:
\begin{quote}
    \textit{``I think that `language' is a bad way to determine what's toxic. You can write terrible things in cordial proper language, and also be kind in crass harsh language. In many cases I couldn't make something not toxic without changing the entire premise, as people were just trying to be rude no matter what.''}
\end{quote}
They later went on to describe the experimenting they had done: 
\begin{quote}
    \textit{``I tested `I love this motherfucker, I'd take a bullet for him any day of the week' it comes back as 100\% toxic. Saying `I genuinely hope you just don\'t wake up tomorrow' is 2\%. There's clearly a flaw in the system''}. 
\end{quote}
\subsubsection{Takeaways}
These responses showcases some generic takeaways users gained through using \tool{}: 
\begin{enumerate}
    \item Current automated moderation models focus mostly on individual words, not higher level meaning.
    \item Words that are considered toxic are sometimes influenced by dialect and slang. 
\end{enumerate} These findings are consistent with how linguists describe modern NLP behaviour\cite{ettinger2019bert}, showing how \textbf{\tool{} enables nontechnical users to quickly understand heuristically how highly complicated language classification models work in practice}. Furthermore, \tool{} provides a canvas for easy experimentation that enables users to effectively find flaws in the system and generate meaningful specific critiques, empowering users to potentially take action where they otherwise would not be able to. 

\subsection{Evaluation 2: Toxic Comment Editing Comparison}
In our second evaluation, we compare the resulting edited text provided by users in study 1 to analyze the effect that editing with or without \tool{} has on the final comments. This will help us not only understand the effect of \tool{}  on online discourse, but also study, more generally, how users might interact and fine tune their language when using an automated toxicity filter. 

\subsubsection{Methodology}
To do this, we recruited 1,458 participants from Amazon Mechanical Turk and asked them to compare two edited comments---one generated by a participant in study 1 in the \tool{}-enabled condition, and another generated by a participant in study 1 in the \tool{}-disabled condition. These comment pairs were randomly selected from the same prompt for each condition. 
Three different participants were asked to assess each set of: original comment, \tool{}-enabled edit, and \tool{}-disabled edit (anonymized as \textit{Edit 1} and \textit{Edit 2} randomly). 
Participants were asked to rate how much they view each comment version as toxic on a five-point scale, and were also asked how well they perceived each edit preserved the general content and intent of the original comment.

\subsubsection{Results}
\begin{figure}[t]
\centering
    \centering
    \includegraphics[width=0.95\textwidth]{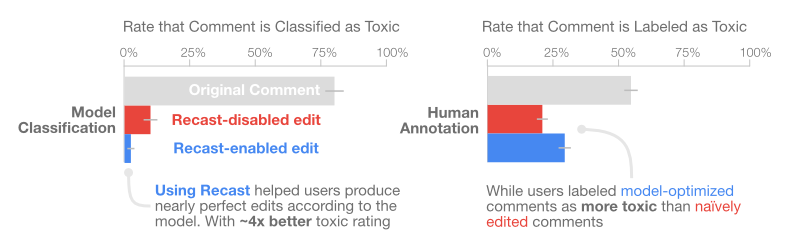}
    \caption{Distribution of toxic labels (either `Agree' or `Strongly Agree' with statement that a comment is toxic for human annotation, or classification by model) for unedited comments, comments edited without \tool{}, and comments edited with \tool{}. Error bars show the 95\% binomial proportion confidence interval under the asymptotic normal approximation. We find that \tool{} does produce optimal comments according to the model. However we find that the model systematically under reports toxicity among edited comments, and that model optimized comments are labeled as on average more likely to be toxic by human annotators.}
    \label{fig:label_hists}
\end{figure}

We found that both the \tool{} enabled and disabled case were statistically indistinguishable with respect to maintaining the original general meaning, both $60\% \pm 2\%$ of the time. \tool{}-enabled and \tool{}-disabled are comparable conditions to look at toxicity, since neither is disproportionately changing the input so as to be incomparable. As expected, the original comments which had already been labeled as toxic, either within the Wikipedia dataset or by the Twitter content filter, were mostly considered toxic, though a significant minority of these original comments were also labeled as not-toxic, highlighting how people perceive toxicity as non-binary. As shown in \autoref{fig:label_hists}, the edits made with \tool{} disabled were generally classified as less toxic than comments made with \tool{} enabled. 

\begin{figure}
    \centering
    \includegraphics[width=0.85\textwidth]{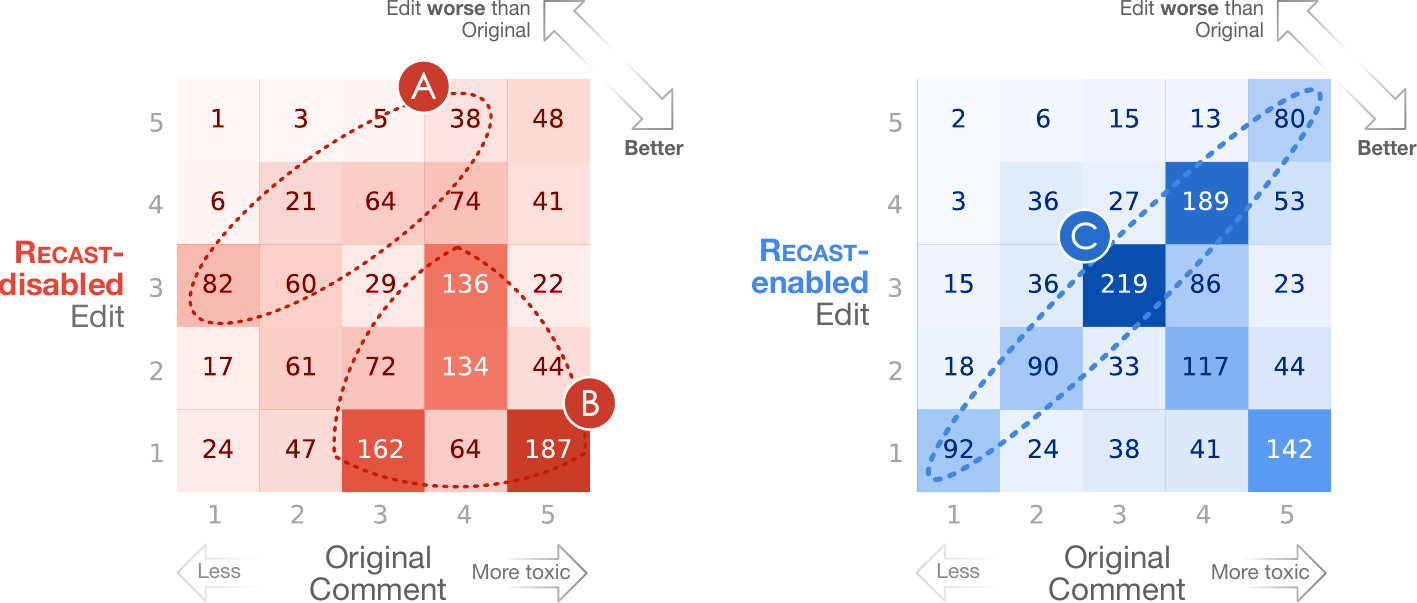}
    \caption{Joint distributions of enabled and disabled edits vs original comment toxicity. Note (A) which showcases the upper diagonal of the disabled case where the resulting toxicity is higher than original toxicity. This region is higher populated than in the enabled case, showing that there is a higher risk of increasing toxicity when writing edits without \tool{}. However (B) showcases that without \tool{}, even high toxicity comments are often reduced. Overall the disabled case shows that without a tool the resultant toxicity is independent of the original. (C) highlights the opposite effect in the enabled case, where the strong representation along the diagonal shows that the resulting toxicity of edits generated using \tool{} is more likely to be similar to the original toxicity, which is a benefit in that there are fewer cases where toxicity is increased in the upper diagonal, but a cost in the lower effectiveness of reducing toxicity in the lower diagonal.}
    \label{fig:label_joints}
\end{figure}

\autoref{fig:label_joints} showcases the joint distributions of the original comment toxicity label in each of the edit conditions. A closer look at the joint distribution suggests that in the \tool{} enabled case, the labeled toxicity is more highly correlated with the original toxicity when compared to the disabled case (Kendall Tau\cite{kendall_1945} of $\tau=0.15$ with $p<0.1$ for enabled and $\tau = -0.03$ with $p>.1$ for the disabled case). This shows that when not using \tool{}, users' ability to write less toxic versions were independent of the original toxicity. However, when using \tool{}, the original toxicity heavily informed the resulting text, where users were less likely to make highly toxic comments much better but much less likely to make comparatively less toxic comments worse.

\subsubsection{Comparing Human Detected Toxicity to Model Classified Toxicity in Edited Comments}
Finally, we looked at the difference between the human annotations of toxicity and model classifications of these edits. We reclassified the outputs of each edit with our fine tuned toxicity classification model, and compared the resulting classifications to the corresponding human labels. As models become more impactful gatekeepers of what language is and is not allowed, any heuristic which works better for the model will be structurally incentivized and potentially become more common. By examining the difference between the moderation using human determined toxicity and by model determined toxicity, we can potentially hypothesize future directions of online discourse as such models become more prevalent. 

In \autoref{fig:label_hists} we see that the \tool{} enabled comments remained classified as toxic $2.4\pm 1.4\%$ of the time, while with \tool{} disabled the model classified the result as toxic $9.9\pm 2.6\%$ of the time. At the same time, edits made using \tool{} do not reliably decrease the human-annotated toxicity of comments, especially for comments with already high toxicity that need the most editing. However, edited comments with originally high toxicity are still classified as toxic by the detection model only a quarter as often as the same comments edited without \tool{}.

\subsubsection{Takeaways}
This analysis shows that:
\begin{enumerate}
    \item  \tool{} was highly effective at helping users reduce toxicity \textit{as detected through the model}, but not as effective at reducing human annotated toxicity.
    \item Therefore when language is optimized for the model (which is what is implicitly incentivized by the deployment of these models), the model ceases to be a good judge of toxicity as determined by human annotators. Language that is less toxic to the model can be more toxic to humans. ``When a measure becomes a target, it ceases to be a good measure''\cite{Strathern1997ImprovingRA}.
\end{enumerate}


\section{Discussion and Implications}

\subsection{\tool{} as a tool for reducing toxicity} 

When discussing the use of \tool{} as a tool for reducing toxicity, we have shown that there are two, potentially competing, meanings of the task. There is the underlying notion of toxicity as language with an adverse effect on people, and there is toxicity as the output of the models used to moderate platforms. The development of models like the Perspective API are predicated on the idea that these two concepts are---if not the same---at least asymptotically close as models improve. However, our work has shown that when users have direct access to the toxicity classification models, their optimized language to match the model does \textbf{not} also optimize human labeled toxicity.

\tool{} is clearly effective at allowing users to reduce the model toxicity of their comments, and does this with good ease of use and accessibility, while maintaining the original meaning. \tool{} also provides a path of least resistance between language that is classified as toxic and language not classified as toxic, and then gives users the power to choose how to use that knowledge. This is useful for a variety of people. Users who may not have an strong fluency of English may inadvertently say things considered toxic without knowledge, and \tool{} provides a frictionless way for these users to make changes and learn what is acceptable. 

On the other hand, some users are better informed than the model when identifying toxic language. We have highlighted how implicit definitions of toxicity used by models are fundamentally different than what humans consider toxic; toxicity is only a meaningful concept in so far as it has an effect on people, not computers. As users pointed out in \autoref{sec:open_ended_Responsed}, slang or dialect may be misclassified as toxic. \tool{} allows these users to both circumvent potentially unjust/biased models, and raise awareness of these issues through explained examples.

While \tool{} does not seem to reduce human labeled toxicity due to users optimizing for the model output, this further emphasizes the need for user recourse and model oversight in this space. \tool{} is an initial effort towards this goal. However, we recognize that there is substantial work that needs to be done from the developer side of these platforms---and potentially on the policy side---in order for the recourse that \tool{} provides to be impactful. 

\subsection{\tool{} as a tool for interpreting toxicity models}

\tool{} was also designed to allow exploration and visualization of toxicity models in order to allow users to understand them and provide a degree of transparency. A large majority of study participants in our evaluation reported that \tool{} helped them understand the toxicity model (\autoref{tab:user_evals}), and many users were able to produce insightful comments about the patterns presented by the model through \tool{}. 
This is a significant component of recourse that \tool{} helps facilitate. When toxicity classification models are used to police language, then users are subject to often biased rules that they may not even know are being applied. Providing actionable recourse is predicated on those affected by those systems understanding \emph{how} they are affected, which is nearly impossible when the system is a black box algorithm. \tool{} hopes to break into that black box and allow end-users and non-experts to then see these rules as they are applied. 

\subsection{Future of Online Discourse}

As machine learning systems are more often deployed to manage moderation online, users will be required to tune their language to match the standards set forth by these models. This will occur regardless of the availability of tools like \tool{}, as only language meeting these models' standards will be visible or filtered (creating a form of survivorship bias). By introducing \tool{} as a way to more directly optimize for these models, we can study the long term evolutionary effect of misapplied filters on future online discourse. Our evaluation quantitatively highlights how the standards required to succeed with a model diverge from human perceived toxicity; \textbf{as model based standards become more prevalent, toxicity according, to human standards, may in fact increase}. This is a troubling trend in our large scale quantitative evaluation that warrants further study. 

\subsection{Limitations and Future Work}
The design of the \tool{} interface is also meant to be generalizable as models improve. Based on prior research in the toxicity detection space, \tool{} utilized the state-of-the-art BERT model to estimate the degree of toxicity in messages and suggest alternative wordings. Despite using a specific type of toxicity detection model, \tool{} is agnostic to BERT specifically and can be easily combined with other machine learning models. Similarly, though the alternative wording suggestion component currently relies on Word2Vec, it serves as a generic framework and is compatible with other embeddings or techniques to generate as broad a space of options for users as possible. \tool{} is also agnostic to model explanation techniques, provided explanations are based on individual words or phrases within a text. Because of our justification in subsection \ref{toxicity-attn-just}, we expect gradient-based model explanations to yield similar results due to flagging of similar tokens. Finally, \tool{} will enable future to work validate the effectiveness of new explainability techniques on model assisted intervention for content moderation. 

With respect to toxicity detection models themselves, our work highlights the numerous challenges associated with automated moderation systems. For example, hateful content may be expressed in multiple ways, e.g., sarcasm, irony, coded text. Users may even use hateful words or phrases to refer to themselves. Instead of investigating different forms of hateful content, we work with a large-scale benchmark corpus with a pre-defined set of toxicity labels. \tool{} can be further extended to handle various formats of toxic speech. Furthermore, \tool{} could allows users to interact with models to take various definitions of toxicity to their limits by optimizing their language. As such, \tool{} may provide a useful backbone for the study of different notions of toxicity. 

Our evaluation of \tool{} was mainly conducted on Amazon Mechanical Turk with annotators in a lab-like environment. As a result, we could not assess the long-term effect introduced by \tool{}. Future work could build upon our research to further investigate whether users will be likely to use tools like \tool{} in their daily interactions on different online platforms, and how \tool{}'s involvement affects users' subsequent participation. 

This also relates to another area of future work: building out implementations of \tool{} that may run in the browser. We actively made decisions to ensure \tool{} is light weight and can be run without significant computational resources on the front-end. By making the code open source, we open an avenue for future work to expand the functionality of \tool{} into a browser extension. This would help both validate the results of the studies we have run by embedding \tool{} in a more realistic scenario, but also make \tool{} accessible to the users who may benefit from it. 

However, we caution future researchers to carefully weigh the ethical implications of widely deploying \tool{}, as such functionality may be highly useful for those users working in good faith, but potentially harmful if used by bad actors. These risks are inherent in any functionality helping users navigate these systems, as explained in \autoref{sec:ethics}. Our study finds that there is an important distinction between two notions of toxicity: (1) Language detected by a model as toxic, and (2) Language that has adverse effects on real people. An inherent risk of visual analytics tools is their ability to only optimize for (1), which while we may hope better aligns with (2) in the future; we find that currently it does not. Thus while \tool{} is effective at helping users acting in good faith to reduce (1), its inability to consistently reduce (2) elucidates flaws in current NLP models rather than the specific design of \tool{}. Finally, an important limitation of any tool is that it requires users to \textit{want} to lower toxicity; which of course is often not the case. However, explicitly handling malicious users is outside the scope of this work, and future work studying when users act maliciously and how to better design human-AI interfaces to \textbf{de-escalate} toxic behaviour before suggesting alternatives---may yield better systems for automated content moderation . 

\section{Conclusion}

In this work, we identified some key problems with the proliferation of automatic toxicity detection models, and 
introduced an interactive tool, \tool{}, to address them. \tool{} provides users the ability to interact with toxicity detection models and visualize how they work. Through these interactions, users are able to make actionable changes to their language in order to reduce toxicity, while gaining generalizable insights about toxicity models. Through multiple large scale user evaluations, we showed the effectiveness of \tool{} in helping users edit text to decrease model defined toxicity while providing interpretable explanations to users. At the same time we highlighted the pitfalls of using toxicity detection models for moderation, as toxicity defined by the model differed from human labeled toxicity. 
We hope \tool{} can help users overcome the challenges of automatic moderation, and further the study of methods to empower \textbf{users} in navigating platforms increasingly governed by machine learning systems.

\section*{Acknowledgements}
We would like to thank anonymous reviewers for their helpful comments. We also thank Dr. Joseph Seering for their feedback on this work. This work was supported in part by NSF grants IIS-1563816, CNS-1704701; DARPA (HR00112030001); gifts from Facebook, Intel, NVIDIA, Bosch, Google, Symantec, Yahoo! Labs, eBay, Amazon.

\bibliographystyle{ACM-Reference-Format}
\bibliography{refs.bib}

\received{October 2020}
\received[revised]{January 2021}
\received[accepted]{January 2021} 

\newpage
\clearpage

\appendix

\section{Appendix}

\subsection{Selected Toxic Comments from Perspective API Dataset}
The following comments were used as the representative comments from the Perspective API Kaggle dataset that users edited in our evaluations. In addition we included a short description to help users contextualize these comments. However these contexts are artificial as the dataset itself does not contain any actual context information.
\begin{enumerate}
    \item 
    \textbf{Context}
    Consider you are an experienced editor on wikipedia, and two new users are attempting to change some edits you have made that you know are correct, however they persist. You are exhausted and lash out, perhaps as this real reply:
    
	\textbf{Comment}
	“Would you both shut up, you don't run wikipedia, especially a stupid kid.”

    \item
    \textbf{Context}
Consider again as an experienced editor you see some incorrect edits attempting to change a historical page about an obscure medieval battle. They appear to have even the most basic facts about the true battle wrong, and frustrated may reply:
	
	\textbf{Comment}
“Hey why you are spreading misconceptions and trying to spread false information to the people. You not even know who won the battle and who lost. you are the cheapest person on the earth. Bloody ignorant fool. Check sources before spreading rumors”

    \item
    \textbf{Context}
In the process of using the website there are a myriad of technical malfunctions preventing you from doing important work, for which you may have a time constraint and as a result may post in a support thread:
	
	\textbf{Comment}
	“HELLO HOW DO I GET SOMEBODY TO FIX SHIT AROUND HERE?”

    \item
    \textbf{Context}
While editing a piece of minutiae on a highly visited page, there is a spirited debate on whether to include an additional sentence which may be seen as redundant to some in explaining the overall point but to you seems necessary, however every time you add this sentence another user removes it, in the discussion thread in reply to this user you may post:
	
	\textbf{Comment}
	“It is not redundant. You are redundant.”

    \item
    \textbf{Context}
Finally in a thread with a fellow editor, who is having similar difficulties as you, you attempt to commiserate with the post:
	
	\textbf{Comment}
	'I know how frustrated you are right now. Stupidity in this place has no limits. Someone with a brain cell or two will eventually show up and clean this mess. Meanwhile hang in there.'

\end{enumerate}

\subsection{Selected Toxic Comments from Twitter}
The following comments were used within the user studies to have examples from outside of the Kaggle dataset. We found these by finding the top posts on the most trending topics on twitter in May, 2020, focusing in particular on Sports and Politics related topics known for high levels of toxicity, and found replies that had been automatically flagged by Twitter's own filtering system. Because we had access to the real context of these replies we also included the full thread when relevant and described the basic context of the tweet. When displaying a thread the final reply is the one users were given to edit.

\begin{enumerate}

    \item
    \textbf{Context}
In this thread discussing the fatal shooting of Ahmaud Arbery, and the delay in finding suspects. User A suggests racial bias while User B disagrees, in this thread the final reply was filtered as toxic and so please edit it from the perspective of User B.

\textbf{User A}
Why did it take too long to arrest two culprits? Cuz they are white and the victim is black and this is America. So yea ... so infuriated

\textbf{User B}
It’s actually because of the Georgia laws but go on with your fake outcry. 

\textbf{User A}
"gEoRgiA LaWs" ...as if 2 black men would get away with murdering a white man thats minding his business. There was clearly no Stand Your Ground defense. Outright execution

\textbf{User B}
I guarantee you didn’t do any research before opening your ignorant trap. But go off girl you definitely doing it

    \item
    \textbf{Context}
After the announcement of the matchups of the NFL regular season for next year, fan of the Dallas Cowboys, User A, talks trash about their rival the Philadeplhia Eagles, when a fan of the Eagles, User B attempts to reply but is filtered:

\textbf{User A}
 2 easy wins against the Eagles

\textbf{User B}
	Keep dreaming scumbag

    \item
    \textbf{Context}

As Liverpool Football Club announced their intention to terminate their contract with player Loris Karius, who is replaced by a different player, Adrián. User A discusses his view that he liked the play of Karius better, while User B disagrees in a very typical British English Dialect that is filtered as potentially toxic.  

\textbf{User A}
I can’t be the only one who rates Karius over Adrian?

\textbf{User B}
Lol what rubbish ! Yes I think you’re the only one mate

    \item
    \textbf{Context}

		Elon Musk, billionaire CEO of electric car company Tesla is suing the county where a Tesla factory is located due to local COVID-19 restrictions preventing the factory reopening. User A is supportive of this move, while User B is critical of it and their response is filtered.
		
\textbf{Musk}
		Tesla is filing a lawsuit against Alameda County immediately. The unelected \& ignorant “Interim Health Officer” of Alameda is acting contrary to the Governor, the President, our Constitutional freedoms \& just plain common sense!
		
\textbf{User A}
		Anything we can do to help? Does reaching out to politicians help in any way?
		
\textbf{Musk}
		Yes
		
\textbf{User B}
Can you get over yourself for five fucking seconds? I'm a fan of your pursuit but you have been a fucking douchebag during this whole pandemic. Go invent something useful, like an actual engine that could get us to Mars in a reasonable amount of time.

\end{enumerate}

\subsection{Study 1}

\begin{figure}[h]
    \centering
    \includegraphics[width=0.75\textwidth]{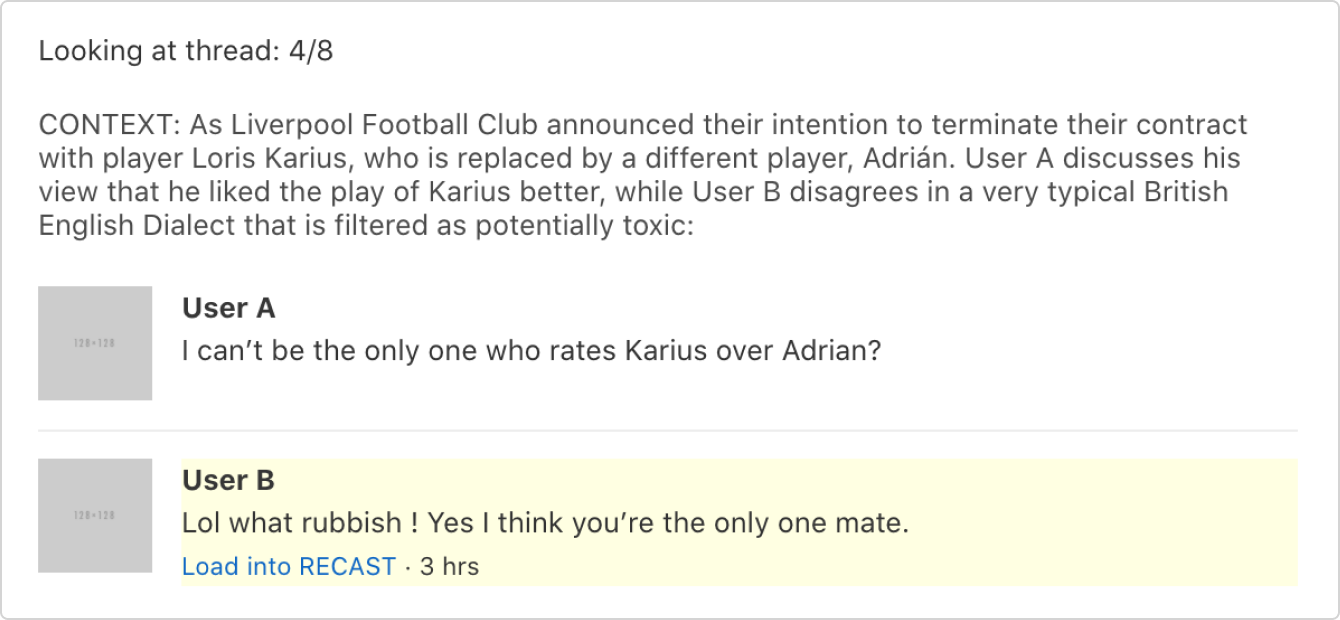}
    \caption{Example of thread provided in study 1. There is a brief explanation of context, followed by a visualization of the comment thread culminating in the comment they are meant to edit. This comment then can easily be loaded into the editing environment.}
    \label{fig:example_thread}
\end{figure}

\autoref{fig:example_thread} shows an example thread that was taken from Twitter. Users were presented with four such threads either from Twitter or Wikipedia before being given access to \tool{}, and then given another four threads after. The order of threads and which comments were provided in each phase of the study were randomized.

\subsection{Study 2}
\begin{figure}[h]
    \centering
    \includegraphics[width=\textwidth]{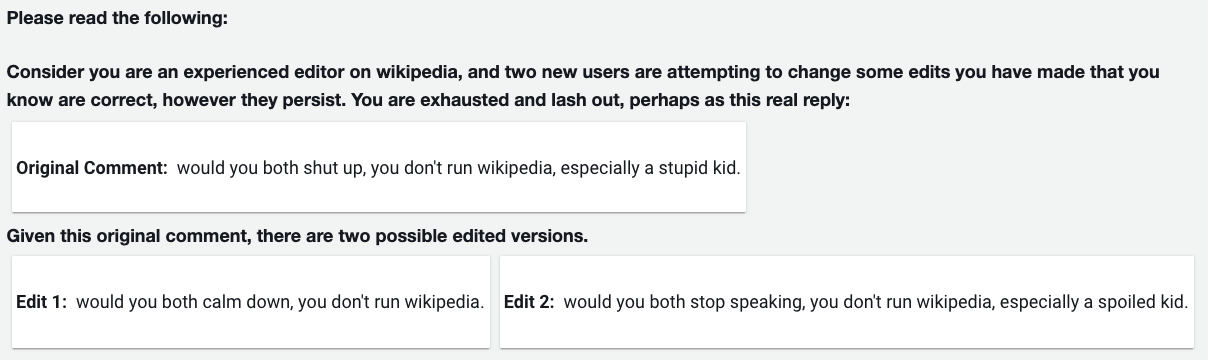}
    \caption{Labeling task in Evaluation 2. Participants are presented the same context provided to participants in study 1, as well as the original comment, its \tool{}-enabled Edit and \tool{}-disabled edit (anonymized as \textit{edit 1} and \textit{Edit 2}). They are then asked to rate the edits as well as the original comment on toxicity, and the edits on their consistency in meaning with the original.}
    \label{fig:labeling}
\end{figure}

\autoref{fig:labeling} shows the user interface for the labelling task. Users were provided the same context as users in study 1, and then given an original comment and then two edited versions (corresponding to either an edit with or without \tool{}) and then asked to rate the toxicity of each edit as well as how well they preserve the meaning. 

\end{document}